%% file: template.tex
\documentclass{sigchi}

\usepackage{balance}  
\usepackage{graphics} 
\usepackage{times}    
\usepackage{url}      
\usepackage{graphicx}
\usepackage{enumerate}
\usepackage{appendix}

\toappear{
This is the authors' version of the work. It is posted here for your personal use. Not
for redistribution. The definitive Version of Record was published in:
CHI '20, April 25--30, 2020, Honolulu, HI, USA.
Copyright is held by the owner/author(s). Publication rights licensed to ACM. \\
ACM 978-1-4503-6708-0/20/04\ldots \$15.00 \\
DOI: \url{http://dx.doi.org/10.1145/3313831.3376327}
}

\usepackage{balance}       
\usepackage{graphics}      
\usepackage[T1]{fontenc}   
\usepackage{txfonts}
\usepackage{mathptmx}
\usepackage{hyperref}
\usepackage{color}
\usepackage{booktabs}
\usepackage{textcomp}

\usepackage{microtype}        
\usepackage{ccicons}          

\usepackage{todonotes}

\newcommand{\superscript}[1]{\ensuremath{^{\textrm{#1}}}}
\def\iVADER{\superscript{1}}
\def\IPD{\superscript{2}}
\def\IBM{\superscript{3}}

\def\plaintitle{GUIComp: A GUI Design Assistant \\with Real-Time, Multi-Faceted Feedback}

\def\emptyauthor{}
\def\plainkeywords{Authors' choice; of terms; separated; by
  semicolons; include commas, within terms only; this section is required.}

\makeatletter
\def\url@leostyle{%
  \@ifundefined{selectfont}{
    \def\UrlFont{\sf}
  }{
    \def\UrlFont{\small\bf\ttfamily}
  }}
\makeatother
\def\pprw{8.5in}
\def\pprh{11in}

\definecolor{linkColor}{RGB}{6,125,233}
\hypersetup{%
  pdftitle={\plaintitle},
  pdfauthor={\emptyauthor},
  pdfkeywords={\plainkeywords},
  pdfdisplaydoctitle=true, 
  bookmarksnumbered,
  pdfstartview={FitH},
  colorlinks,
  citecolor=black,
  filecolor=black,
  linkcolor=black,
  urlcolor=linkColor,
  breaklinks=true,
  hypertexnames=false
  }

\begin{document}

\title{\plaintitle}

\numberofauthors{1}
\author{%
  \alignauthor{Chunggi Lee\iVADER, Sanghoon Kim\iVADER, Dongyun Han\iVADER, Hongjun Yang\iVADER,\\ Young-Woo Park\IPD, Bum Chul Kwon\IBM, Sungahn Ko\iVADER\thanks{The corresponding author} \\
    \affaddr{\iVADER Electrical and Computer Engineering, UNIST, \{cglee, seiker, handy113, hj42, sako\}@unist.ac.kr }\\
    \affaddr{\IPD Graduate School of Creative Design Engineering, UNIST, ywpark@unist.ac.kr}\\
    \email{\IBM IBM Research, bumchul.kwon@us.ibm.com}}\\
}

\maketitle

\begin{abstract}

Users may face challenges while designing graphical user interfaces, due to a lack of relevant experience and guidance. 
This paper aims to investigate the issues that users with no experience face during the design process, and how to resolve them. 
To this end, we conducted semi-structured interviews, based on which we built a GUI prototyping assistance tool called GUIComp. 
This tool can be connected to GUI design software as an extension, and it provides real-time, multi-faceted feedback on a user's current design. 
Additionally, we conducted two user studies, in which we asked participants to create mobile GUIs with or without GUIComp, and requested online workers to assess the created GUIs. 
The experimental results show that GUIComp facilitated iterative design and the participants with GUIComp had better a user experience and produced more acceptable designs than those who did not.

\end{abstract}




\input{1introduction}

\input{2relatedwork}

\input{3preliminarystudy}

\input{4design}
\input{5experiment}

\input{6result}

\input{7discussion}

\input{8conclusion}
\input{9acknowledgement}

\balance{}

\bibliographystyle{SIGCHI-Reference-Format}
\bibliography{sample}

\end{document}

%% file: 1introduction.tex
\section{Introduction}
Different people may use contrasting criteria when evaluating designs, which makes design tasks challenging. 
One suggested approach to effectively produce acceptable designs is to perform iterations among design stages~\cite{Nielsen93}: produce many shareable design alternatives~\cite{Dow11, Kulkarni14}, compare the alternatives~\cite{Tohidi06}, and evaluate how general users feel while employing the designs (e.g., attention~\cite{Weinschenk11}). 
Iterative design also helps people improve their design skills for quickly prototyping alternatives and allows training time with many design trials to achieve the desired goals (e.g., simplicity, theme expressions, visual aesthetics, and creativity). 

Though performing iterative design is effective, doing so may not be easy for novices.
By \textit{novices}, we refer to people who have difficulty in performing the iterative design, have little confidence in making design decisions, and follow ``trial and error'' as their design strategy~\cite{Ahmed03} due to lack of or little design experience.
Examples include students who need to produce designs for a class or developers who need to not only program, but also design Graphical User Interfaces (GUIs) for the program (e.g., freelance or independent app developers~\cite{Codewithchris}).
We conjecture that they encounter obstacles whenever they iterate the design stages. 
For example, beginning with a blank canvas in the initial prototyping stage can often be overwhelming for the novices due to the difficulty in conceptualizing designs and harmonizing the concepts within the given design constraints (e.g., layouts or color themes)~\cite{Deininger17}.
Novices are likely to make mistakes through design iterations, a few of which may result in the end in design failure.
Generally speaking, it is hard for novices to recognize mistakes in advance.  
As such, there is a need for an end-to-end system that can assist novices by lowering the barriers in the design process.

In this work, we aim to design a tool for assisting novices.
To achieve this goal, we conducted semi-structured, in-depth interviews with 16 participants to understand the difficulties of mobile GUI design. 
Based on the difficulties observed during the interviews, we designed a tool, \textbf{GUICompanion (GUIComp)} by integrating three different types of feedback mechanisms: \textit{recommendation, evaluation, and attention}. 
We designed GUIComp as a browser add-on, so that it can be easily linked to existing GUI prototyping tools.
In our experiment, we linked GUIComp with a base tool, called Kakao Oven~\cite{KakaoOven} and asked 30 participants to design GUIs for user profile input and a list of products with either GUIComp or the base tool. 
Then we asked 47 Amazon Mechanical Turk (AMT) workers to assess the resulting designs.
The results indicate that the proposed tool helped users who lacked experience in mobile GUI design to easily begin and develop a GUI design within a short period.
With GUIComp, the users were able to efficiently start their designing with examples, check whether their design seemed acceptable, and produced it as they intended guided by visual complexity scores. 
The results also show that mobile GUIs produced with GUIComp were more acceptable designs to general users than those produced with the base tool. 
The participants reported that designing mobile GUIs with GUIComp was more enjoyable, satisfactory, and affordable than with the base tool.
We believe that our approach to providing real-time multi-faceted feedback can be applicable to non-mobile GUI design tasks (e.g., web design).

The contributions of this work include the 1) characterization of the difficulties that users encounter while designing GUIs by conducting semi-structured interviews, 
2) design and evaluation of GUIComp, an end-to-end system that integrates multi-faceted feedback for mobile GUI design and facilitates an iterative design process, and 3) lessons learned from the study and design guidelines for GUI prototyping assistance.

%% file: 2relatedwork.tex
\section{Related work}

\subsection{GUI Prototyping Tools}
Many tools have been developed for GUI prototyping. 
Adobe Photoshop and Illustrator are popular among designers for providing many toolboxes for drawing various simple and complex shapes~\cite{Photoshop}. 
Several tools have been proposed to allow easy prototyping by reducing the time for interaction and adding animation. 
Adobe XD~\cite{AdobeXD}, InVision~\cite{InVision}, Sketch~\cite{Sketch}, and Principle~\cite{Principle} are tools in this category.
Other tools have also been proposed to allow more rapid prototyping, such as UXPin~\cite{UXPin}, Proto.io~\cite{Protoio18}, and Axure~\cite{Axure}, all of which are equipped with a large number of ready-made elements. 
Despite the helpfulness of these tools in rapid prototyping, they may not be sufficient for users, because they do not provide timely feedback on the users' designs to improve quality.

\subsection{Approaches for Assisting GUI Prototyping}
Many approaches exist for assisting GUI prototyping, and they can be categorized into three groups. 
The first approach allows users to browse examples. 
Existing studies report that browsing examples inspires designers and leads to alternative designs~\cite{Herring09, Dow10,Lee10a}. 
For example, ``d.tour'' proposed by Ritchie et al.~\cite{Ritchie11} allows users to search for examples by using user-provided keywords. 
The second approach enables users to interactively explore, extract, and use specific design elements from the examples.
The extracted elements can be directly used as a template or building block in prototyping. 
Tsai and Chen's framework~\cite{Tsai07} is an early system for supporting template-based mobile user interface design.
However, in this approach, users are limited in terms of which elements they can extract, such as structures, layouts, and styles.
Rewire~\cite{Swearngin18} automatically converts a UI screenshot to a vector representation of design elements, which users can import into their graphic editing tools and revise for their own purposes. 
The third approach aims to partially or fully automate the design stages using computational methods. 
O'Donovan et al.~\cite{Donovan14, Donovan15} develop DesignScape which provides users with various layout suggestions for refinement and brainstorming.
Todi et al.~\cite{Todi18, Todi16} propose an approach that produces familiar designs by restructuring existing layouts and an interactive layout sketching tool, which offers real-time design optimization with suggestions of local and global changes to improve the usability and aesthetics. 
In contrast, ReDraw~\cite{Moran18c} and pix2code~\cite{Beltramelli17} are designed to produce a code, which can be executed to create an application, generated by a convolutional neural network model based on a sample UI image. 
Moran et al.~\cite{Moran18} propose an approach that automatically checks whether the users' design implementation follows a design guideline provided as an image.
Despite their usefulness, automated approaches have several weaknesses.
First, users are often excluded from the automatic design recommendation process, except for the choice of its input (e.g., an image of an example UI).
In addition, an automated approach does not guarantee the high quality of its output; the quality is often delegated to users.
Automated methods can also prevent users from freely envisioning creative designs constrained by their output examples, as reported in previous studies~\cite{Oh18, Donovan14}.
These weaknesses can be overcome by supporting users to lead the design process, but rare studies exist on how to support.
We investigate how people can lead design efforts, assisted by the machine acting as a smart companion during the process.

\subsection{Metrics for Measuring Visual Complexity}

Metrics for evaluating web pages and mobile GUIs~\cite{Ivory01, Reinecke13, Koch16, Oulasvirta18} already exist.
Web page evaluation metrics focus on the visual complexity of websites with individual and numeric factors (e.g., word count or graphics count~\cite{Ivory01}), whereas those for mobile GUIs concern detailed elements that contribute to the overall visual complexity of designs, such as words, color, image counts and sizes, symmetry, balance~\cite{Ivory01, Reinecke13}, and layouts~\cite{Koch16}.
There are other approaches for evaluating visual complexity of mobile GUIs~\cite{Riegler18, Riegler15, Miniukovich14b, Ines17}.
Miniukovich and Angeli~\cite{Miniukovich14b} propose metrics for measuring visual impressions of mobile GUIs and demonstrated that their model can explain 40\% of the variation of subjective visual complexity scores. 
Their experiment also revealed that dominant colors and the symmetry of UIs are correlated with aesthetics, while color depth information is more correlated with visual complexity. 
The model is further extended to measure the visual complexity of desktop GUIs~\cite{Miniukovich15}.
Riegler and Holzmann~\cite{Riegler18} propose eight visual complexity metrics for evaluating mobile GUIs.
Their metrics evaluate the quality of the number of UI elements, misalignment, imbalance, density, element smallness, inconsistency, color, and typographic complexity. 
The metrics were mathematically formulated and evaluated through a user study.
In this work, we use Riegler and Holzmann's metrics~\cite{Riegler18}, because they can present visual complexity in numeric values in real-time.

%% file: 3preliminarystudy.tex
\section{Identifying Difficulties Encountered During GUI Prototyping}

We conducted semi-structured interviews to understand the difficulties users face during GUI prototyping.
We chose the mobile GUI domain, because mobile GUI design is a difficult design task due to innate characteristics, such as small screens with touch-based input~\cite{Khalid15, Zhang05, Doosti18, Wu19, Miniukovich15}.

\textbf{Participants: } 
We recruited 16 participants (3 women, avg. age: 23.56) at a university in South Korea, who were interested in designing GUIs for mobile applications. 
We recruited the novices only by noting in the advertisement that individuals with any previous GUI design experience were not eligible to participate.
The recruited participants were all undergraduate students with various engineering majors, namely, electrical, computer, mechanical, and biomedical engineering. 
No participants reported any experience with designing mobile GUIs. 

\textbf{Procedure: }
As the participants entered the experimental room located in a university building, they were first asked to fill out a demographic questionnaire that sought information about their GUI design experience. 
Each participant was then given a computer (Intel i7 and 16GB RAM with a 2560 $\times$ 1440 resolution, 27-inch monitor) that had a GUI design tool (Kakao Oven~\cite{KakaoOven}). 
We chose Kakao Oven as the GUI prototype software for three reasons. 
First, we looked for a design tool for the novices that provided pre-defined icons (e.g., pins) and design components (e.g., buttons), which would prevent the additional difficulty of making the components from scratch during the design. 
Second, we wanted a tool with the description and instructions in Korean so that participants could easily learn how to use it. 
Third, we sought a tool that was similar to other popular tools used outside South Korea so that the study's findings could be generalized to other tools (e.g., Sketch~\cite{Sketch}, UXPin~\cite{UXPin}). 
As Kakao Oven fit the criteria well among the many alternatives available, we chose to use it for our study. 
\autoref{fig_system_overview}~A is a screenshot of Kakao Oven, presenting an example design for shoe sales.
The Canvas panel (A2) indicates the area where the users create GUI designs. 
The Element panel (A3) includes various pre-built design elements (e.g., text, button, and tooltip) users can easily customize and apply to their design on the Canvas panel. 
We used an eye-tracker (Eyelink 1000+, sampling rate up to 2kHz) to analyze users' behavior.

The participants were given as much time as they needed to become familiar with the tool during the practice session. 
The supervising researcher also answered any questions they had about the tool during this session; subsequently, we conducted the eye-tracker calibration. 
Next, we introduced the task that the participants had to perform, which was to create a GUI for an online shopping mall application that lists product information along with product images. 
Note that we did not impose detailed or strict design requirements, such as the color theme or product domain. 
Accordingly, we proposed an open-ended task to investigate the obstacles encountered by the users while they planned a design and building a GUI from scratch using the given tool. 
The resulting list of identified obstacles served as a guide in deriving the design requirements for the proposed design assistant tool. 
The task session for each participant lasted about an hour, and the participants were filmed during the session.
After completing the task, each participant watched the recorded video along with the researcher and discussed the challenges that the participant encountered during the GUI design process. 
We transcribed the audio conversations from these discussions.

\subsection{Identifying Difficulties Encountered by Participants}
We coded the transcribed audio conversations using the grounded theory approach~\cite{Muller14}, to identify roadblocks encountered by participants~\cite{kwon_visual_2011}. 
First, two of the authors of this paper individually coded the reports on the encountered difficulties and then, reviewed and discussed the collected codes to identify common themes. 
Afterward, we were able to group the participants' difficulties into three distinct categories: (1) choosing the design direction, (2) measuring the design quality, and (3) determining where the viewer's attention would fall. 
After identifying these categories, we coded each participant's response relating to the difficulties as either ``0'' or ``1'' for each category. 
Only one category was assigned to each difficulty (for example, a difficulty could be coded as [0,0,1]). 
Finally, we collected and compared the results.
When there were disagreements, the coders resolved them through a discussion. 
If the coders did not reach agreement, another author of this work resolved the disagreement.
The inter-coder agreement level was 85\% as computed by the Pearson correlation coefficient. 
Based on the observations, we characterized and illustrated the three main difficulties that users may face during GUI prototyping.

\textbf{D1: OK, how do I start?: } 
One common observation was that the participants initially struggled to determine how to go about the process. 
They often began the task by creating a few GUI elements (e.g., text, buttons, and images) by dragging and dropping from the element panel to the canvas area. 
However, soon after, the participants  deleted their created elements and began looking for examples online. 
This series of actions demonstrated that the participants encountered difficulties in the initial stages. 
Twelve participants reported that they could not easily conceptualize their prototyping direction~\cite{Nielsen92b} and sought examples for inspiration~\cite{Chang12, Herring09, Kumar11, Swearngin18} and quality comparison~\cite{Tohidi06}.
Eight participants reported that they found it difficult to start designing GUIs without concrete examples. 
Twelve participants commented that they needed a template that provided a skeleton layout with areas where elements could be replaced based on the user's judgment.

\textbf{D2: Is my design OK?: }
The second difficulty was that they were uncertain whether they were making good or bad design choices. 
One participant reported facing difficulty in deciding the optimal size of the elements (e.g., icons, images, and buttons) for the application users. 
Eight participants mentioned that it was a challenge to choose colors that would be aesthetically pleasing to potential customers.
During the task session, one participant looked online for GUI evaluation methods or guidelines, although the returned results were not helpful; the participant reported two reasons for this: the searched guidelines were often too abstract (e.g., \textit{``use conventional elements''}) or the advice was difficult to apply to her current design context (e.g., regarding the instruction to \textit{``use about 44 squared pixels for a touch interface,''} she did not have 44 squared pixels to allocate).

\textbf{D3: Will users see what I want them to see?: }
The third difficulty was related to determining where the viewers' attention would fall in the process of modifying the design (D3). 
Four participants stated that they had no clue how to determine which areas would be mostly viewed by users. 
Based on our observation, this difficulty seemed to lead to frequent changes in the position and size of the images.
A participant reported, \textit{``It was hard to guess which part the customer will look for first, as all of the text and images are important, I think.''}
Another participant stated that \textit{``I want to emphasize selling points in my design, but I do not know which part is most proper to be the point.''}
When we asked the participant what made her think about this emphasis, she answered that she had experience using an online shopping mall where important information, such as coupons or product prices were not sufficiently evident to shoppers on a mobile screen.

\subsection{Requirements of a Tool for Assisting GUI Prototyping}
Based on the identified difficulties, we derived the requirements for a tool that can solve these problems while assisting GUI prototyping:
\begin{enumerate}[R1:] 
    \item Provide examples to guide users in how to begin and modify the design,
    \item Evaluate the current state of the design,
    \item Indicate the projected areas that users will see,
    \item Ensure R1--R3 on the most up-to-date design in progress,
    \item Ensure that R1--R3 are met non-intrusively, and
    \item Meet R1--R3 using an add-on that can be attached to a web-based prototyping tool.
\end{enumerate}

We found that providing timely feedback is essential to solve problems.
Feedback can be considered one of the most powerful instruments  for helping users achieve a desired goal~\cite{Hattie07}, in particular for creative work~\cite{Sadler89}. 
Feedback given during an early design stage can allow users to iteratively improve the quality of the design~\cite{Dow10, Kulkarni14}.
Thus, we derived three requirements (R1, R2, R3) that aim to provide relevant examples (R1),  evaluation of the design (R2), and areas of interest for target users (R3).
In addition to the three requirements directly determined from the three difficulties (D1, D2, and D3), we included another three requirements.
We added R4, because it is critical for users to receive prompt feedback on their current designs, as early and timely feedback could improve creative work~\cite{Kulkarni15, Kulkarni14}.
R4 also implies that the design process should not involve an offline computation so that the iterative processes of designing, reviewing, and revisiting can be efficiently supported~\cite{Buxton10}.
In addition, we included R5, because we did not want users to be interrupted by such feedback while making their design choices. 
R5 implies the need for a dedicated space where feedback can be visually displayed.
The purpose of R6 is the easy installation and compatibility of GUIComp.  
Following these requirements, we designed a tool for assisting GUI designers, as discussed in the next section.

%% file: 4design.tex
\section{GUIComp: An Add-on for GUI Design Guidance}
We introduce a web-based add-on tool, called GUICompanion (GUIComp), which provides prompt feedback on GUI designs. 
GUIComp is designed as an add-on for mobile GUI design applications so that it can be easily adapted to any other tool with proper configuration of client-server communication.

\begin{figure*}[t]
\centering
\includegraphics[width=0.85\textwidth]{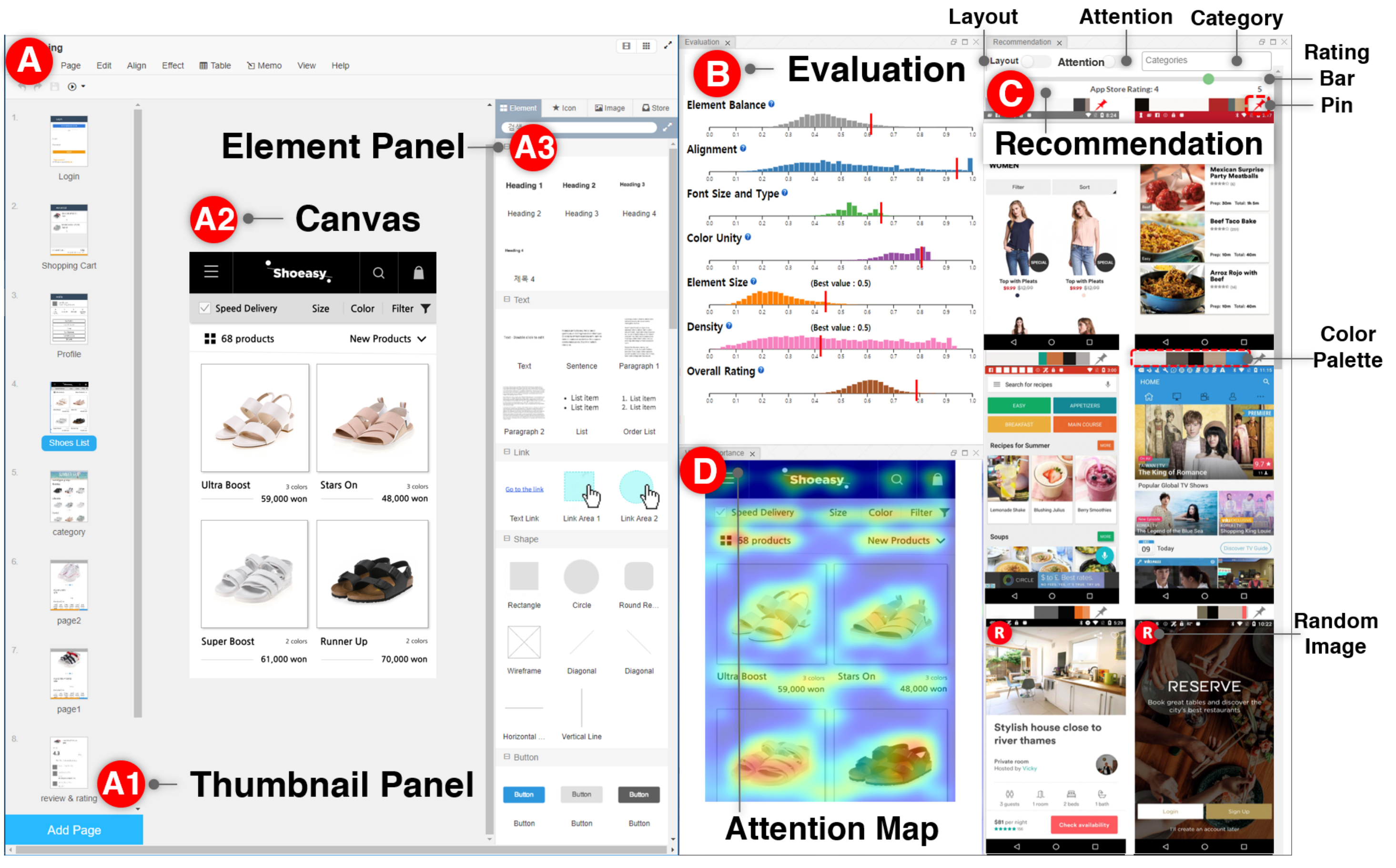}
\caption{Panels (A1-A3) are interfaces of an online GUI prototyping tool (the base tool). Three feedback three feedback panels in the GUIComp (Add-on)--Evaluation (B), Recommendation (C), and Attention Panel (D). 
}
\label{fig_system_overview}
\vspace{-0.3cm}
\end{figure*}

\subsection{Tool Overview}
GUIComp consists of three panels (the evaluation, recommendation, and attention panels), where each panel fulfills the three requirements (i.e., R1--R3) described in the previous section.
We implemented the tool in the Chrome browser as an extension~\cite{ChromeExtension} that uses responsive web technologies to capture and process information in real time (R4), and provides feedback in a separate window outside the user's design canvas (R5). 
As an extension, GUIComp can be linked to any online GUI prototyping tool (R6).
For demonstration and evaluation purposes, we linked GUIComp to Kakao Oven~\cite{KakaoOven}, a base GUI prototype tool. 
GUIComp captures the most up-to-date design states by using open source libraries (i.e., MutationObserver), generates feedback (i.e., visual complexity evaluation), recommendation templates, and attention heatmaps based on the states, and presents them on the three panels (\autoref{fig_system_overview}~B--D).
In the following sections, we describe how we captured the users' design states, parsed the web elements, and extracted the design elements (e.g., element types, dimensions, and color maps) and we subsequently introduce the panels, metrics, and datasets used for GUIComp.

\subsection{Capturing and Processing Users' Design on Canvas}
GUIComp captures user design states and returns two different outputs, HTML elements and images.
Then, using the two outputs as inputs, GUIComp generates three different types of feedback (examples, evaluation scores, and attention heatmaps).
When there is an interaction on the user canvas (\autoref{fig_system_overview}~A2), an internal HTML file for describing the user canvas is also updated. 
Once an update is detected by using MutationObserver, a JavaScript API, GUIComp converts the HTML file into a json file using the html2json~\cite{html2json} and html2canvas~\cite{html2canvas} libraries. 
GUIComp saves the json file temporarily in the Chrome extension's local storage to prevent Cross-site Request Forgery (CSRF) vulnerabilities.
Then, GUIComp sends the json file to the processing server along with an image of the GUI on the canvas.

The processing server extracts GUI components' ID, type (e.g., button), and attributes (e.g., color, width, and height) from the json files and computes i) the dominant colors using OpenCV~\cite{OpenCV} and ii) the visual complexity scores using the equations proposed by Riegler and Holzmann~\cite{Riegler18}, which are described further in the Evaluation Panel section.
GUIComp also captures and stores the GUI image from the canvas; see \autoref{fig_system_overview} A2.
The captured GUI image is used as an input for the FCN-16 model, and the model provides an attention map score as an output.

\subsection{Recommendation Panel}

\subsubsection{Description of In-the-Wild Mobile GUI Data (RICO)}
To create a pool for recommendation templates, we considered both the ERICA~\cite{Deka16} and RICO datasets~\cite{Deka17a}, but chose to use RICO, as it includes one of the most extensive lists of mobile application screenshots. 
The RICO dataset~\cite{Deka17a} contains 72,219 mobile application GUI screenshots, meta-data, and element hierarchy information from 9,700 Google Play Store popular apps.
The dataset also contains meta-data of the applications, including the average rating, the number of installs and downloads, and the category assigned for the app marketplace. 
The average ratings and categories of the apps are utilized for filtering examples in \autoref{fig_system_overview} (C).
We manually reviewed each screenshot and excluded those that were inappropriate for use in GUIComp. 
Specifically, we excluded screenshots of playing games, commercial advertisements, pop-up menus, Android basic screens (e.g., the home screen), data loading, password typing, black images, web views, and maps.
In the end, we used 6683 screenshots from 2448 applications for this work.
\vspace{-0.15cm}

\subsubsection{Generating Recommendations from HTML}
From the HTML file, the GUI structure information (e.g., TextView, EditText, Button, ImageView, and ImageButton) is also extracted and used as an input for Stacked Autoencoder (SAE) and the k-nearest neighbor algorithm, as suggested by Deka et al.~\cite{Deka17a} for recommending examples that are similar to the current GUI on the canvas.
In our implementation, the encoder has 13,500 (90$\times$50$\times$3) input (the image size is the same ratio as the RICO dataset) and 64 output dimensions with two hidden layers of 2048 dimensions and 256 dimensions with the ReLU activation function~\cite{Nair10}.
The decoder has the same but reversed architecture of the encoder. 
For training and recommendation, we used the RICO dataset. 
We used 90\% of the data (6154 images) for training and kept the rest for validation.
We used the Adadelta optimizer and mean squared error as the loss function.
During the training, the validation loss was stabilized at 4.96 after about 1900 epochs for 1.5h with 512 batches. 
Using the trained model, we generated a 64-dimensional representation per GUI screenshot for the RICO dataset.
The k-nearest neighbor algorithm was implemented with the brute search algorithm and the cosine distance metric.

\subsubsection{Presenting Similar and Random Examples}
The recommendation panel provides users with example templates that are relevant to and inspirational for their design goals.
It is not feasible to present all examples from the RICO dataset to users and ask them to explore the list without any guidance.
Furthermore, as users make progress in their designs, they might need different examples for different purposes.
Thus, our goal was to provide a list of the most relevant examples at the moment based on the current state of the design.
To achieve this goal, we computed similarity scores between the user's design and examples so that we could rank the examples in the order of similarity.

In addition to similar examples, we decided to present ``random (four images)'' examples for two reasons. 
First, giving random examples increases diversity in design choices, as the standard affinity algorithm can generate recommendation lists that are similar~\cite{Hurley11}.
Second, users can rethink the choices they made (e.g., the layout).
By viewing intriguing alternatives, users may have a chance to think outside the box.
To distinguish random examples from recommended examples, we show a visual mark (\raisebox{-0.3ex}{\includegraphics[width=0.03\linewidth]{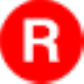}}), as shown in \autoref{fig_system_overview} C (bottom). 
When there is no component on the user canvas (e.g., at the beginning), the panel is populated with random examples. 

We speculated that users might want to ``keep'' some templates while changing others as they make progress.
Thus, we allowed users to pin examples so that they maintain the selected examples in the list; they can also unpin the examples (\autoref{fig_system_overview} PIN \raisebox{-0.3ex}{\includegraphics[width=0.04\linewidth]{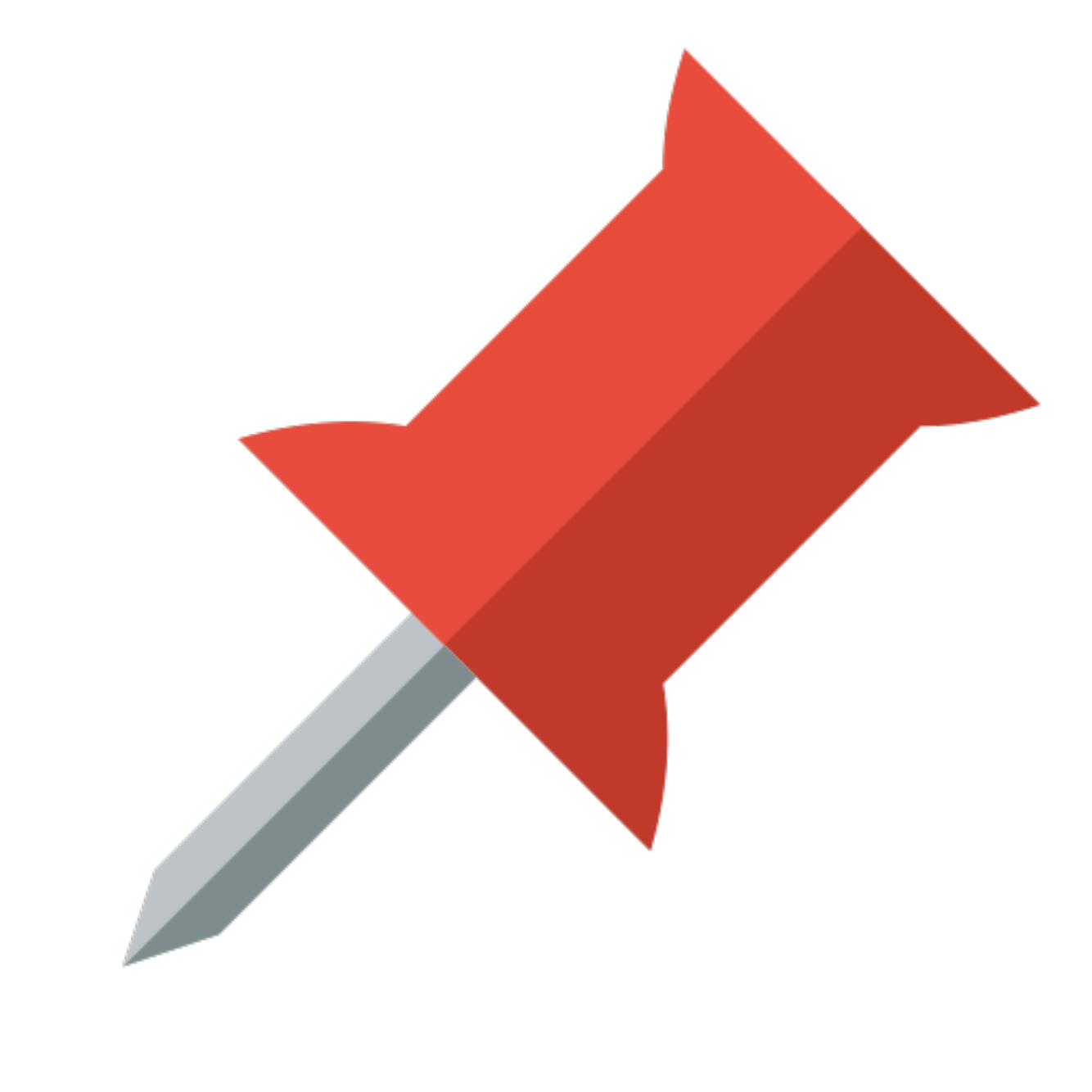}}).
The panel also lists dominant colors in their color palettes (e.g.,~\autoref{fig_system_overview} Color Palette \raisebox{-0.3ex}{\includegraphics[width=0.2\linewidth]{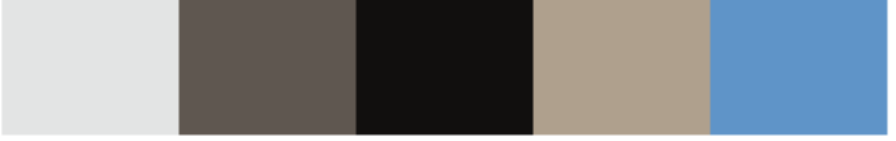}}) generated from the OpenCV library.
When a user hovers over a color in the palette, the RGB information of the color is shown to help users refer to and learn the combinations of colors in the examples.
When a user clicks on an example template, the user canvas is cleared and then elements of the clicked template populate the user canvas in the same layout and alignment as shown in the template.  
To allow this automated element-populating function, we first mapped the information of the RICO leaf-level nodes (e.g., x, y, width, height, and type) to that of the Kakao Oven elements (\autoref{fig_system_overview} A3), and computed the ratio of RICO's nodes to screen width and height and Oven's elements to the canvas width and height. 
Then, the drag--mock library~\cite{DragMock} updates the HTML code to indicate that new elements were added.
Note that users can recover previous designs by using a revert function provided by the base tool. 


\subsection{Attention Panel}

To visually show how much attention will be paid to elements on a GUI design (R3), we incorporated Bylinskii et al.'s attention model~\cite{Bylinskii17}. 
We feed the captured image from the user design on the canvas into the attention model (FCN-16)~\cite{Bylinskii17} that has been pre-trained with a graphics design importance (GDI) dataset~\cite{Donovan14} to produce a heatmap with pixel values ranging from 0 to 255.
\autoref{fig_system_overview} D shows an attention heatmap with the given GUI on the canvas in \autoref{fig_system_overview} A2. 
The color map used in the heatmap linearly ranges over blue-green-yellow-red, where the red indicates the areas that will receive the most attention.
The more intense the red on an element on the heatmap, the higher the importance of the element in the attention map.  
Guided by the heatmap, users can expect where viewers will direct their attention. 
Based on the information, users can revise the design for optimal guidance.
With the attention map, users can balance the expected audience's attention by revising the original design.
For example, when a text field unintentionally receives too much attention, a user may change the position or size of the text to redirect the audience's attention to more important areas. 

\subsection{Evaluation Panel}
\label{sec_panel_eval}

\subsubsection{Generating Visual Complexity Metrics}
To provide feedback on GUI quality (R2), we incorporated visual complexity metrics.
We also expected that, by using the metrics as evaluation rubrics, users could easily discern possible issues in their current design~\cite{Yuan16}. 
A review of previous studies that involved experimental evaluation enabled us to find two candidate metrics~\cite{Riegler18, Miniukovich15}.
Of the two metrics, we chose the one by Riegler et al.~\cite{Riegler18} because it enables real-time evaluation (R4) and allows for an easy-to-understand numerical result presentation. 
From the authors' original seven metrics, we excluded two: the number of UI elements and inconsistency.
We omitted the former because the absolute number of UI elements does not reflect the quality of a design, given that the number varies depending on the design goal. 
We eliminated inconsistency because it applies only when a design has more than one GUI page. 
We also modified the original metric names into more intuitive ones to help users understand the terms. 
Specifically, we changed imbalance to \textbf{element balance}, misalignment to \textbf{alignment}, color complexity to \textbf{color unity}, typographic complexity to \textbf{font size and type unity}, and element smallness to \textbf{element size}.
We did not change the term \textbf{density}.
For metrics that feature 0 as the best score (i.e., element balance, alignment, color utility, and font size and type unity), we reversed the score range from 1 to 0 to 0 to 1 so that 1 became the best score. 
We did not convert the scores for element size and density because the midpoint (i.e., 0.5) is considered the best score for the metrics~\cite{Riegler18}. 
These metrics were also used as rubrics for the evaluation by the online workers. 
We describe the metrics in detail as follows:
\textbf{Element Balance (best score: 1.0)} refers to the overall symmetry, balanced element distribution (e.g., consistent space between elements), and skewness of the elements. 
\textbf{Alignment (best score: 1.0)} pertains to the checking of alignment among elements. During computation, three vertical (left, middle, and right) and three horizontal (top, middle, and horizon) imaginary lines are drawn for each element to measure the score. 
\textbf{Color Unity (best score: 1.0)} shows the color use based on the ratio of dominant to non-dominant colors.
\textbf{Font Size and Type Unity (best score: 1.0)} investigates the consistency of font sizes and types present in the text. 
\textbf{Element size (best score: 0.5)} is intended to verify whether elements are excessively small or large for mobile interfaces. Scores lower than 0.5 mean the elements are small, while scores higher than 0.5 imply the elements are large, on average. 
\textbf{Density (best score: 0.5)} computes how much space is occupied. Scores of less than 0.5 translate into simplicity in design, whereas higher scores imply over-populated designs. 

\subsubsection{Presenting Visual Complexity Scores with User Ratings. }

To help users understand the strengths and weaknesses of their current design compared to those in the RICO dataset (R2)~\cite{Hartmann10}, the evaluation panel (\autoref{fig_system_overview} B) presents six visual complexity scores and one overall rating score for the user's current design.
This panel uses six histograms, each of which shows the distribution of examples in the RICO dataset with its corresponding complexity score on a horizontal scale.
The vertical red bar on each scale shows how high or low the corresponding visual complexity score of the user's design is compared to that of the examples. 
For instance, the example design in the user canvas (\autoref{fig_system_overview} A2) can be evaluated as a high-quality design based on the positions of the red bars over the distributions in each evaluation dimension (\autoref{fig_system_overview} B). 

To support R4, GUIComp computes new scores whenever a user interaction occurs over the design canvas (A2).
For example, when a button element is dragged from the element panel (A3) and dropped to the canvas (A2), the evaluation panel updates with newly computed scores to present the effects of the dropping of the button.
Note that the current design's scores are marked with a thick, vertical, red bar in each histogram (\autoref{fig_system_overview} B).
The scores of the recommended examples are depicted with black bars when a user hovers over an example in the recommendation panel.


\subsection{Implementation Notes}
We used two servers: one (Intel Xeon E5-2630, 2.40GHz, 128GB RAM) for processing the captured user interaction and the other (Intel Xeon E5-2630, 2.20GHz, 128GB RAM, 2 $\times$ Tesla P100-PCIe-12GB) for model training and prediction. 
GUIComp uses several web development libraries, including Django~\cite{Django}, Keras~\cite{Keras} D3.js~\cite{d3js}, and Bootstrap~\cite{Bootstrap}.
Note that in the performance experiment, when a user interaction happens on the base tool's canvas, it takes 725, 885, and 750 ms for each panel to produce feedback using the data processing server. 
We also logged all user interactions, GUI states, and attributes for qualitative analysis. 

%% file: 5experiment.tex
\section{User Study Design}
To evaluate GUIComp, we conducted a user study, in which 30 participants were asked to create two GUI designs: one with restrictions and one without. 
Then we presented the designs to online workers to evaluate the quality of the designs. 
This evaluation was guided by three primary research questions:
\begin{enumerate}[RQ1:]
    \item Does GUIComp help users make better designs according to general users than Kakao Oven alone?
    \item Does GUIComp provide users with a more fun, fulfilling, and satisfactory experience?
    \item Do users perform iterative design process to overcome the difficulties during prototyping?
\end{enumerate}

\subsection{Participants, Procedure, Apparatus, and Tasks}
We recruited 30 participants with an advertisement at a university according to the definition of the novices in the introduction section.
As participants entered the experiment room, they were asked to fill out a form on demographic and background information, including name, age, gender, major, prototyping, and development experience and time.
Then, the participants were randomly assigned to either the Control Group (CG) or the Experiment Group (EG).
The CG participants were allowed to use only a base GUI prototyping tool, called Kakao Oven (\autoref{fig_system_overview} A only), whereas the EG participants were given the proposed tool and the base tool (\autoref{fig_system_overview} A--D). 
We used Kakao Oven for the same reason as in the previous participatory study. 
We also concerned that if we used a different tool as the basis and found strange or disappointing results, we would not know if those results were due to an unsuccessful design or simply to some un-analyzed differences between the two tools.
The experiment was a \textbf{between-subject} study to minimize learning effects and was carried out with a computer (Intel i7, 3.40GHz, 16GB RAM, 2560x1440 27-inch monitor). 

At the beginning, the experimenters explained the purposes and goals of the experiment to both groups for 5 minutes. 
Then the participants in both groups watched the explanatory video for 5 minutes (using Kakao Oven) or 15 minutes (Kakao Oven and GUIComp, including the visual metrics explanation). 
We also provided concise descriptions of the visual metrics in a popup window, which the participants could mouse over to read during the study.
The participants were allowed as much time as possible to familiarize themselves with the tool assigned to them, and explore the tool's features.
In addition, we told participants that 10\% of them would receive an incentive (US\$10), in addition to the base payment (US\$20) based on the perceived quality scores rated afterward by online workers.
Finally, the study began after the experimenters calibrated the eye-tracker (Eyelink 1000+, sampling rate up to 2 kHz) for the gaze transition analysis. 

During the experiment, we gave the participants two tasks: a situation with design restrictions given by clients (e.g., embedding brands~\cite{Wu19}) and one without any restrictions.
The two tasks we used were a user profile interface with restrictions (T1) and an item-listing interface without any restrictions (T2). 
We chose the two interfaces as the tasks for users with little design experience, because we thought that 1) the difficulty level was appropriate, as the basic design components are provided by Kakao Oven, such as icons and buttons, and 2) the interfaces are commonly requested, given that many apps require input of user information and show items in a listing interface, regardless of the app category.
Inspired by Lee et al.'s persona-based web page design~\cite{Lee10a}, we used in T1 a persona-based GUI design with detailed restrictions.
There were no other restrictions, which means the participants were allowed to customize their UIs as much as they desired in order to win the incentive. 

Next, we describe the personal-based task used in the experiment as follows:
\textit{A user's name is Elaine (email address: elaine@gmail.com). Elaine has 12 notifications. There is one shipping item and two items in the shopping basket. Elaine left 31 reviews.}
Goal: Design a profile user interface for Elaine. 
Tasks and restrictions:
1) Your GUI should include basic information about Elaine. 
2) You can add detailed information (e.g., order list, delivery tracking, cancellation returns, discount coupon, Service center, profile management).
3) Use the headings, text, shape, button, and pagination components of the given tool in your basic design layout.
4) Enter textual information about Elaine.
5) Choose colors for each component (optional).
6) Choose the font and font size for each text component (optional).


\begin{figure*}[t]
\centering
\includegraphics[width=0.8\linewidth]{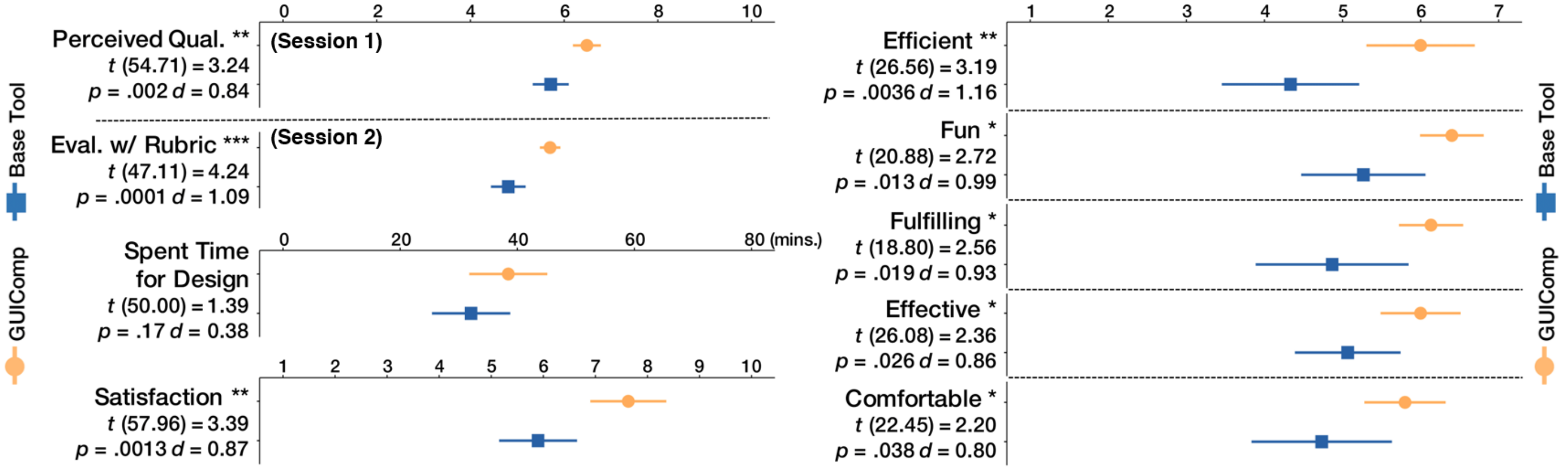}
\vspace{-0.1cm}
\caption{The online workers' ratings of the participants' designs (left) and exit survey results (right). The dot and the whisker represent the mean and the 95\% confidence interval, respectively. 
}
\label{fig_quality_time_result}
\label{fig_ux_result}
\vspace{-0.3cm}
\end{figure*}
After completing the tasks, the participants responded to an exit survey about usability and user experience with GUIComp (RQ2), with the survey comprising items on efficiency, fulfillment, effectiveness, ease of use, ease of learning, user-friendliness, consistency, fun, and satisfaction~\cite{Lund01, Albert13}. 
These items were rated by the participants on a 1--10 scale (1: ``highly disagree," 10: ``highly agree") for satisfaction scores, and on a seven-point Likert scale (1: ``highly disagree," 7: ``highly agree") for the other scores. 
The users also provided preference scores for each feedback panel (1: ``least preferred," 7: ``most preferred").

Overall, we recruited 32 participants, but two could not complete the experiment due to sudden system failures. 
Thus, 30 participants remained (18 men and 12 women, avg. age: 22.4, all from engineering schools, 15 users/group) and they had little or no experience (i.e., novices) but were interested in GUI design (less than a month: 6 participants, 1-3 months: 1, 3-6 months: 1, no experience: 22).
4 participants experienced a prototyping tool as their personal hobby (2: Kakao Oven, 2: Adobe Photoshop).
None of the other participants had previous prototyping tool experience. 
We excluded the eye tracking data for 6 participants in the EG, as the data did not contain full prototyping processes due to a malfunction.  
Thus, we used 18 sets of eye tracker data (9 participants $\times$ 2 tasks) in our eye tracking-based analysis.
We analyzed EG groups' gaze data only, because we are interested in how GUIComp helps users (RQ1--RQ3).

%% file: 6result.tex
\section{Experiment Results, Analysis, and Implications}
Here, we report our evaluation approach with online workers. 
As the goal of GUIComp is to help users produce designs that are acceptable to general users, such relative acceptance levels can be measured by the scores given by online workers, who can be regarded as general mobile GUI users. 

We evaluated the designs in two sessions. 
In Session 1, we asked 26 MTurk workers to evaluate the perceived design quality on a 0-10 scale.
No rubric was provided in Session 1 to collect general users' subjective assessment scores. 
In Session 2, we asked 21 other MTurk workers to evaluate the designs with a rubric, which were the evaluation metrics (\autoref{fig_system_overview} B-element balance, alignment, color unity, font and element size, and density), as the assessment criteria.
We gave the workers the rubric to provide minimum guidance for preventing inconsistent and subjective evaluation~\cite{Yuan16}. 
We also provided detailed descriptions for each metric to help the workers understand and use the metrics for evaluation. 
We believe that using the rubric was appropriate for guiding the evaluation, because it includes the indexes for visual complexity (e.g., color, size, density, and alignment), which affect the overall quality of the GUI designs~\cite{Doosti18, Wu19, Miniukovich15}.
Next, we present the evaluation results using the Welch's t-test and the Kruskal-Wallis test, due to unequal variance in the data. 



\subsection{GUIComp Users Produce Designs Acceptable to General Users (RQ1)}
In this section, we present statistical test results for the evaluation scores.
In doing so, we merged the assessment scores of both tasks in each session, and ran Welch's t-tests to see whether GUIComp was effective regardless of the task type.
Note that we report the results with 95\% confidence interval plots, as shown in \autoref{fig_quality_time_result}, where the dot and the whisker represent the mean and the 95\% confidence intervals, respectively. 
\autoref{fig_quality_time_result} (left, top row, ``Perceived Qual.") indicates that the GUIs produced with GUIComp were considered more acceptable designs (i.e., higher scores) by general users than those produced with Kakao Oven ($t$[54.71]=3.24, $p=$0.002, Cohen's $d$=0.84).
We observe the same result for Session 2 results (\autoref{fig_quality_time_result}, left, second row, ``Eval. w/ Rubric"), where the GUIs were evaluated with the rubric ($t$[47.11]=4.24, $p<$0.001, Cohen's $d$=1.09). 
However there is no statistically significant difference in the time spent performing the design tasks in the two sessions ($t$[50.00]=1.39, $p=$0.17).
Overall, the results indicate that using GUIComp helps people, especially those who are in the beginner stage of designing GUIs or have never designed GUIs, produce designs acceptable to the general users, without spending any additional time on the designs. 

\begin{figure*}[t]
\centering
\includegraphics[width=\linewidth]{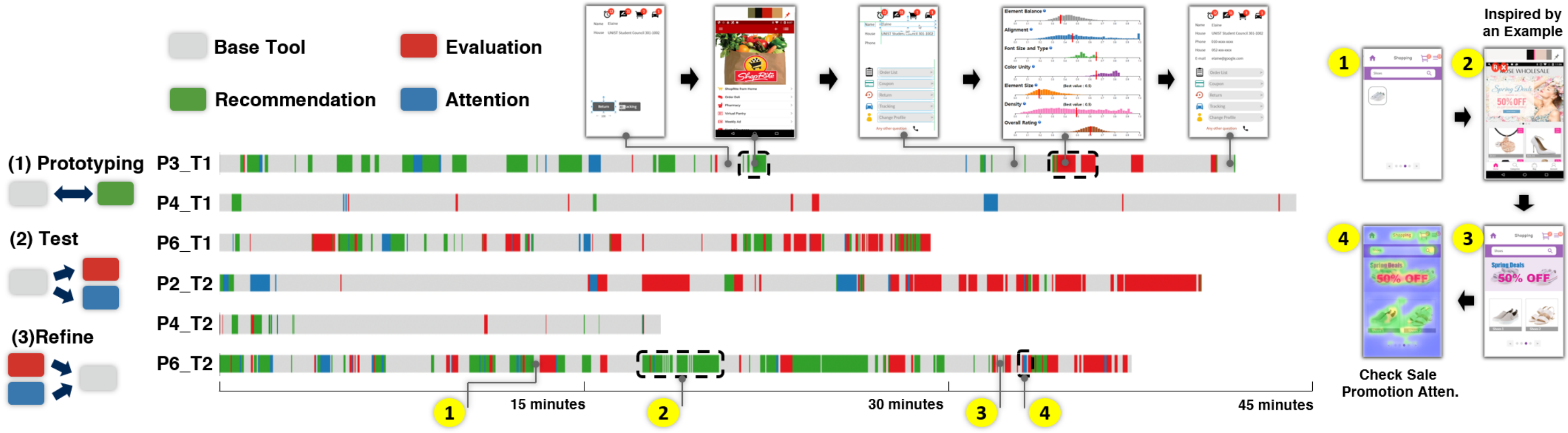}
\vspace{-0.5cm}
\caption{User gaze sequence data on the feedback panels. 
}
\label{fig_view_sequence}
\vspace{-0.5cm}
\end{figure*}

\subsection{Using GUIComp Is Enjoyable and Satisfactory (RQ2)}
\autoref{fig_quality_time_result} (right) presents the exit survey results: The participants who used GUIComp felt that using the tool was more \textbf{efficient} ($t$[26.56]=3.19, $p$=0.003, Cohen's $d$=1.16) and \textbf{effective} ($t$[26.08]=2.36, $p$=0.026, Cohen's $d$=0.86) in GUI prototyping than those who did not use it.
Participants who used GUIComp while prototyping also had more \textbf{fun} ($t$[20.88]=2.72, $p$=0.013, Cohen's $d$=0.99), and felt more \textbf{comfortable} ($t$[22.45]=2.2, $p$=0.038, Cohen's $d$=0.8) and \textbf{fulfilled} ($t$[18.8]=2.56, $p$=0.019, Cohen's $d$=0.93), than those who did not use the tool.
Finally, participants who used GUIComp were also statistically significantly more \textbf{satisfied} with the tool ($t$[57.96]=3.39, $p$=0.0013, Cohen's $d$=0.87) than those who did not use GUIComp. 

In the post-experiment interviews, we observed multiple reasons for the positive results, including the feedback provided in real-time, which enabled efficient iteration of their design process with fun and high satisfaction. 
Participants who used GUIComp liked the feedback feature and enjoyed the way their multi-faceted feedback (R1) was dynamically updated in real-time (R4) throughout the design process.
Participant 2 (P2) said \textit{``It was nice to design with the data (i.e., feedback provided by GUIComp), as up to now, my intuition has been the only option that I can rely on during design.''}

None of the participants reported any perceived clutter or distraction caused by the real-time feedback from GUIComp during the design process.
Instead, the real-time feedback (R4) was a feature of interest to some of the participants.
P1 stated that \textit{``I think it was fun seeing that feedback was changed in real-time based on my interactions.''}
We do not see statistically significant differences between the two tasks regarding how user-friendly and how easy to use and learn the tools were. 
The results show that GUIComp could be added to existing design technologies without sacrificing their usability.

\subsection{Users Employ Multi-faceted Feedback for Overcoming Difficulties in the Iterative Design Process (RQ3)}
In this section, we present the results of the eye-tracking data analysis, where we find how GUIComp facilitated the iterative designs, and how the users responded to the feedback during their design iterations. 
\autoref{fig_view_sequence} presents the panel gaze sequences of six participants who used GUIComp (three sequences per task, which received the highest quality scores, were chosen) based on participants' eye movements between panels.
We also present example user designs to describe how the designs evolved, based on the feedback from GUIComp.

Note that we denote eye gaze locations on the panels in four different colors--gray for the base tool (\autoref{fig_system_overview} A), red for the evaluation panel (\autoref{fig_system_overview} B), green for the recommendation panel (\autoref{fig_system_overview} C), and blue for the attention panel (\autoref{fig_system_overview} D). 
In addition, we use the terms ``prototyping'', ``testing'', and ``refining'' borrowed from Nielsen's iterative design~\cite{Nielsen93} to capture distinctive design stages with GUIComp. 

We also consider the intentions of each panel at a high level.
By prototyping, we mean the transitions between the base tool and the recommendation panel (\raisebox{-0.6ex}{\includegraphics[width=0.3\linewidth]{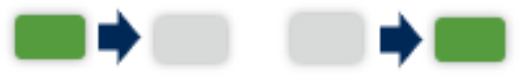}}). 
By testing, we mean the transitions from the base tool to the evaluation or attention panel (\raisebox{-0.6ex}{\includegraphics[width=0.3\linewidth]{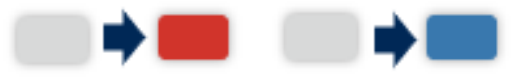}}). 
Lastly, we define refining as the transitions from the evaluation or recommendation panel to the base tool (\raisebox{-0.6ex}{\includegraphics[width=0.3\linewidth]{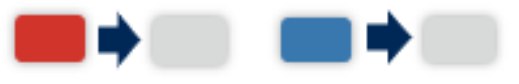}}).

First, we observe frequent color changes in each user's gaze visualization in \autoref{fig_view_sequence}, which mean that the participants performed an iterative design process, forming many different transition patterns.
All six participants used the recommendation panel (green) frequently in the early stage, and some participants used the recommendation panel until the middle stages during their design, which is consistent with what Kulkarni et al.~\cite{Kulkarni14} found in their study.
Based on our reviews on the interview transcripts, we conclude that participants viewed examples of the recommendation panel for inspiration at the beginning of the prototyping, and mainly for solving D1.
P9 stated how useful GUIComp was during the early design stage for solving D1: \textit{``this tool is useful, in particular in the early design procedure with the abundant feedback and references.''}
P2 described her design strategy for solving D1 during prototyping as follows: \textit{``I started with the recommendation panel in the early stage to search as many templates as possible so that I could work with many good options.''}
Other participants, such as P9, also shared their experiences of using examples for solving D1:  \textit{``The recommendation panel was useful early on... With the many references, I could decide which layout to use and start with."}
We observe actual patterns of this strategy in \autoref{fig_view_sequence} (e.g., P3\_T1 and P6\_T2 middle green parts).
When examples are used, users tend to gradually build their design by parts, referring to the different parts in the examples. 
This behavior is interesting, because we expected users to employ the design strategy of taking all the elements of the best example in their opinion and submitting their resulting design by slightly modifying the example. 

We observe that participants tended to make more use of the evaluation panel in the later design processes (e.g., \autoref{fig_view_sequence} P3\_T1, red parts at the end).  
Participants' gaze often moved to the canvas to find an element to enhance the design quality after viewing  the evaluation panel. 
We speculate that participants wanted to confirm quality of their intermediate or final design quality by using visual complexity scores (testing and refining) in an effort to solve D2. 
This effort was illustrated by P13, who said, \textit{``I consistently checked if the alignment is correct, colors are properly used, and the element balance is good. 
In this way, I have created a GUI that I like.''} 
Other participants, such as P14, expressed a similar opinion as P13: \textit{``Mostly, I paid attention to increase the evaluation scores during my prototyping and had a high quality design.''}

During the prototyping, the participants monitored the attention information, to balance viewers' focus to solve D3.
P6 directly expressed her intention of using this view--\textit{``I can see which parts will attract the viewers on which I could edit my design.''} 
It is of interest that P6 checked whether viewers' attention would lie on the most important sales words in the design (\autoref{fig_view_sequence} P6-T2 `50\% OFF'), producing a high-quality design. 
Although most participants enjoyed the attention view, \textit{``I like the (attention) panel, because it is very intuitive and easy to understand,''} as P2 stated, one participant indicated a concern that using this view may not always be easy, since it often requires interpretations of the results. 


To understand which type of feedback played a critical role in improving the participants' designs, we reviewed the user preference average scores (recommendation: 5.0, evaluation: 4.93, attention: 5.53)~\cite{Pu11}.
We ran Kruskal-Wallis ANOVA tests, which indicate that there is no significant difference in user preference among the three types of feedback (\textit{p}=0.60). 
In terms of the time spent on each panel for reviewing feedback (avg. duration--recommendation: 168s, evaluation: 126s, and attention: 27s), we find significant differences among the panels (\textit{p}=0.001, \textit{F}=7.40), according to the ANOVA test. 
The Tukey post-hoc analysis results indicate that users spent more time reviewing the feedback from the recommendation and evaluation panels than that from the attention panel. 
We do not find any statistically significant difference in the eye-gazing duration between the recommendation and evaluation panels. 
To sum up, we think that all of the feedback in the panels contributed to improving user satisfaction in using GUIComp. 
In terms of contribution to improving designs, we conjecture that evaluation and recommendation feedback could have played more important roles, according to the time that users spent reviewing and reflecting on the feedback for their designs.

%% file: 7discussion.tex
\section{Lessons, Limitations, and Discussion}
This section describes lessons we learned from this study and discusses limitations of our work.

\textbf{Provide Explanation for Feedback or Allow Users to Intervene in the Feedback Generation Process: }
We believe design assistance tools can further support users by providing additional feedback for ``why" (justified feedback~\cite{Ngoon18}) to help users better understand the reason for the given feedback. 
In this study, participants requested justified feedback in GUIComp when the feedback was not clear.  
For example, a few participants often wanted to know exactly which areas or elements contributed the most negatively to the evaluation score.
P6 struggled to find areas for improvement, and stated, \textit{``I designed in a way to increase my alignment score, but the score did not go up as I intended."}
The same issue occurred with the recommendations.
When we asked how we could improve GUIComp, P5 commented, \textit{``There were many recommended examples, but I could not understand why some examples were recommended to me."}
Users wanted to understand the reasons for the scores and the recommendations.

One solution is to provide a relevant explanation for the feedback.
For example, when a user does not understand a low score in alignment, an assistance tool can point out the misaligned elements, so users can take appropriate actions.
Another solution is to include users in the automated process. 
For example, users can be invited to intervene in the example recommendation process.
After viewing the initial suggestions, users may provide feedback by setting constraints on desired designs (e.g., background color) or by directly labeling selected examples as ``good" or ``bad."
We can consider providing more granular feedback using semantics of UIs~\cite{Liu18}, by referring to what each visual element means and how it is supposed to be used, in order to help users better interpret the results.
Future research may investigate how to make the system's feedback more interpretable and how to incorporate users' feedback in the system's feedback. 

\textbf{Conflict between User Thought and Feedback: }
In this study, we provided participants with three types of feedback. 
The participants often struggled with feedback that was opposite to what they believed. 
In particular, some participants were reluctant to accept numeric scores in the evaluation panel. 
In the experiment, P14 intentionally changed the layout so that some elements were overlapping. 
This action decreased the evaluation scores. 
After the experiment, P14 stated, \textit{``Though I saw my scores dropped as I made the element overlap, [...] I still think my (overlapping elements) design looks better than others with high scores.''}
The strength of numeric scores is the clear interpretability.
However, when the scores do not match what users believe, the scores may not be appreciated. 
In addition, the evaluation scores, which do not reflect subjective elements, layouts, or themes, may be disregarded.
Future research should investigate how to reconcile the potential conflicts either by augmenting the numeric score with more convincing, descriptive, textual feedback or by improving the feedback using users' input.

\textbf{Automation and Feedback: }
As we discussed above, the participants often demanded more explanations and explicit guidance. 
In this study, we provided feedback in a non-intrusive manner so that the users could be able to contemplate the feedback for themselves. 
In this way, we believe that users can lead their design efforts using their creativity~\cite{Oh18}. 
However, the participants revealed that they would have appreciated some automated correction of details, such as misalignment. 
Future research should investigate the effects of such automated designs or corrections on the users' experience, and quality of their work. 
For instance, an automated GUI and code generation technology from a GUI screenshot is one of the emerging technologies in GUI prototyping, including pix2code~\cite{Beltramelli17}, ReDraw~\cite{Moran18c}, and DesignScape~\cite{Donovan14, Donovan15}.
However, other studies report that auto-generation can take the imagination of developers out of the design process and remove opportunities for learning how to improve designs~\cite{Donovan14,Oh18}.
In addition, there is a wide design space between automation and manual design. 
Future studies should investigate how to integrate automated support, while facilitating user-driven, creative design processes for GUIs.

\subsection{Limitations and Future Work}
As our goal was not ranking the features, we could not report report superiority and interplay effects among the features in this work. 
As our experiment was with the novices and Kakao Oven, we could not tell the effectiveness of our approach with experienced users and other baseline tools. 
We plan to conduct an experiment for comparing and ranking the features and for investigating the effect of our approach with more experienced users and other tools.

%% file: 8conclusion.tex
\section{Conclusion}
We propose GUIComp, a feedback-based GUI prototyping tool to assist users, based on semi-structured interviews to identify the roadblocks that users face during the design process. 
GUIComp provides three types of feedback (evaluation, attention, and recommendation), which are dynamically updated as users continuously make changes in their design. 
The experimental results indicate that GUIComp allows users to create more acceptable mobile GUI designs, while providing a more fun, fulfilling, and satisfactory experience. 

%% file: 9acknowledgement.tex
\section{ACKNOWLEDGMENTS}
We thank Michael Muller and Jeffrey Nichols for their helpful comments and suggestions. 
This work was supported by the National Research Foundation (NRF) grant (No. 2017R1C1B1002586), ITRC (Information Technology Research Center) support program (IITP-2020-2017-0-01635), and the Institute of Information\&communications Technology Planning\&Evaluation (IITP) grant (No.2017-0-00692, Transport-aware streaming Technique Enabling Utra Low-Latency AR/VR Services), funded by the Korea government (MSIT).